\documentclass[12pt,nofootinbib,tightenlines]{revtex4} 
\usepackage{amsmath}
\begin{document}
\def\vs{\vskip .2in} 

\centerline {\Large\bf Decoherent Histories Quantum Mechanics and}
\centerline{\Large \bf Copenhagen Quantum Mechanics\footnote{\sl The conclusion from the authors' paper  {\it Quasiclassical Coarse Graining and Thermodynamic Entropy}, {\sl Phys. Rev. A}, {\bf 76}, 022104 (2007), arXiv:quant-ph/0609190 modestly edited and reformatted. The references have not been updated. Preparation of this version was supported in part by NSF grant PHY18.028105. }}
\vskip .1in
\centerline{\bf\large Murray Gell-Mann and James. B. Hartle\footnote{and Department of Physics, University of California, Santa Barbara, CA}}
\vskip .1in
\centerline{\bf\large Santa Fe Intstitute, Santa Fe, NM}
\vskip .1in 


\vskip .2in

We discuss the  relation between our approach to quantum mechanics, based on coarse-grained decoherent histories of a closed system, and the approximate quantum mechanics of measured subsystems, as in the ``Copenhagen interpretation.''
The latter formulation {\it postulates} (implicitly for most authors or explicitly in the case of Landau and Lifshitz \cite{LL58}) a classical world and a quantum world, with a movable boundary  between the two. Observers and their measuring apparatus make use of the classical world, so that  the results of  a ``measurement'' are  ultimately expressed in one or more ``c-numbers''. 

We have emphasized that this widely taught interpretation, although successful, cannot be the fundamental one because it seems to require a physicist outside the system making measurements (often repeated ones) of it. That would seem to rule out any application to the universe, so that quantum cosmology would be excluded. Also billions of years went by with no physicist in the offing. Are we to believe that quantum mechanics did not apply to those times?

In this discussion, we will concentrate on how the Copenhagen approach fits in with ours as a set of special cases and how the ``classical world'' can be replaced by  a quasiclassical realm. Such a realm is not {\it postulated} but rather is {\it explained} as an emergent feature of the universe characterized by the Hamiltonian $H$, the quantum state $|\Psi\rangle$, and the enormously long sequences of accidents (outcomes of chance events) that constitute the coarse-grained decoherent histories. The material in this paper can be regarded as a discussion of how  quasiclassical realms emerge.

We say that a `measurement situation' exists if some variables (including such quantum-mechanical variables as electron spin) come into high correlation with a quasiclassical realm.  In this connection we have often referred to fission tracks in mica. Fissionable impurities can undergo radioactive decay and produce fission tracks with randomly distributed  definite directions. The tracks are there irrespective of the presence of an ``observer''. It makes no difference if a physicist or other human or a chinchilla or a cockroach looks at the tracks. Decoherence of the alternative tracks induced by interaction with the other variables in the universe is what allows the tracks to exist independent of ``observation'' by an ``observer''. All those other variables are effectively doing the ``observing''. The same is true of the successive positions of the moon in its orbit  not depending on the presence of ``observers'' and for density fluctuations in the early universe existing when there were no observers around to measure them.

The idea of ``collapse of the wave function'' corresponds to the notion of variables coming into high correlation with a quasiclassical realm, with its decoherent histories that give true probabilities. The relevant histories are defined only through the projections that occur in the expressions for these probabilities. Without projections, there are no questions and no probabilities.   In many cases conditional probabilities are of interest.  The collapse of the probabilities  that occurs in their construction is no different from the collapse that occurs at a horse race when a particular horse wins and future probabilities for further races conditioned on that event become relevant.

The so-called ``second law of evolution'',  in which a state is `reduced' by the action of a projection, and the probabilities renormalized to give ones conditioned on that projection, is thus not some mysterious feature of the measurement process.  Rather it is a natural consequence of the quantum mechanics of decoherent histories, dealing with alternatives  much more general than mere measurement outcomes. 

There is thus no actual conflict between the Copenhagen formulation of quantum theory and the more general quantum mechanics of closed systems. Copenhagen quantum theory is an approximation to the more general theory that is appropriate for the special case of measurement situations. Decoherent histories quantum mechanics is rather a {\it generalization} of the usual approximate quantum mechanics of measured subsystems.

In our opinion decoherent histories quantum theory advances our understanding in the following ways among many others: 

\begin{itemize}
\item Decoherent histories quantum mechanics extends the domain of applicability of quantum theory to histories of features of the universe irrespective of whether they are receiving attention of observers and in particular to histories describing the evolution of the universe in cosmology.  
 
\item The place of classical physics in a quantum universe is correctly understood as a property of a particular class of sets of  decoherent coarse-grained alternative histories --- the quasiclassical realms \cite{GH93a,Har94b}.
In particular, the {\it limits} of a quasiclassical description can be explored. Dechoherence may fail if the graining is too fine. Predictability is limited by quantum noise and by the major branchings that arise from the amplification of quantum phenomena as in a measurement situation. Finally, we cannot expect a quasiclassical description of the universe in its earliest moments where the very geometry of spacetime may be undergoing large quantum fluctuations. 

\item Decoherent histories quantum mechanics provides new connections such as the relation (which has been the subject of this paper)  between the coarse graining characterizing quasiclassical realms and the coarse graining characterizing the  usual thermodynamic entropy of chemistry and physics. 

\item Decoherent histories quantum theory helps with understanding  the Copenhagen approximation. For example, measurement was characterized as an ``irreversible act of amplification'', ``the creation of a record'', or as ``a connection with macroscopic variables''. But these were inevitably imprecise ideas. How much did the entropy have to increase, how long did the record have to last, what exactly was meant by ``macroscopic''? Making these ideas precise was a central problem for a theory in  which measurement is fundamental. But it is less central in a theory where measurements are just special, approximate situations among many others. 
Then characterizations such as those above are not false, but true in an approximation that need not be exactly defined.

\item Irreversibility clearly plays an important role in science as illustrated here by the two famous applications to quantum-mechanical measurement situations and to thermodynamics. It is not an absolute concept but context-dependent like so much else in quantum mechanics and statistical mechanics. It is highly  dependent on coarse graining, as in the case of the document shredding \cite{Gel94}. This was typically carried out in one dimension until the seizure by Iranian ``students'' of the U.S. Embassy in Tehran in 1979, when classified documents were put together and published. Very soon, in many parts of the world, there was a switch to two-dimensional shredding, which still appears to be secure today.
 It would now be labeled as irreversible just as the  one-dimensional one was previously. The shredding and mixing of shreds clearly increased the entropy of the documents, in both cases by an amount dependent on the coarse grainings involved. Irreversibility is evidently not absolute but dependent on the effort or cost involved in reversal.
 
\end{itemize}. 

The founders of quantum mechanics were right
in pointing out that something external to the framework of wave function
and Schr\"odinger equation {\it is} needed to interpret the theory.  But
it is not a postulated classical world to which quantum mechanics does not
apply.  Rather it is the initial condition of the universe that, together
with the action function of the elementary particles and the throws of quantum
dice since the beginning, explains the origin of quasiclassical realm(s)
within quantum theory itself.

\end{document}